\documentclass[a4paper,12pt]{article}
\usepackage{amssymb}
\usepackage{amsmath}
\usepackage{epsfig}
\usepackage{graphicx}
\usepackage{slashed}
\usepackage{xcolor,color}
\usepackage{ulem}
\usepackage{subfigure}
\usepackage[toc,page]{appendix}
\usepackage[makeroom]{cancel}
\usepackage{tikz}
\usetikzlibrary{shapes.geometric,positioning,decorations.pathreplacing,decorations.pathmorphing,patterns,decorations.markings}

\makeatletter

\@addtoreset{equation}{section}
\makeatother
\def\be{\begin{equation}}
\def\ee{\end{equation}}
\def\bea{\begin{eqnarray}}
\def\eea{\end{eqnarray}}
\def\bgn{\begin{align}}
\def\egn{\end{align}}

\def\({\left(}
\def\){\right)}
\def\<{\left<}
\def\>{\right>}

\def\({\left(}
\def\){\right)}
\def\<{\left<}
\def\>{\right>}
\def\!{\right|}
\def\|{\left|}

\def\[{\left[}
\def\]{\right]}

\def\+{\bar}

\def\ng{{\negthinspace}}

\def\bR{{\bf R}}

\def\l{{{\ell}}}

\def\P{{\cal{P}}}
\def\V{{\cal{V}}}

\usepackage{hyperref}
\hypersetup{colorlinks,%
  citecolor=blue,%
  linkcolor=blue,%
  pdftex}

\begin{document}

\begin{titlepage}
\vskip1cm
\begin{flushright}
% UOSTP {\tt 1512001}
\end{flushright}
\vskip0.25cm
\centerline{
\bf \large 
Structure of deformations in Jackiw-Teitelboim black holes with matter
} 
\vskip0.8cm \centerline{ \textsc{
 Dongsu Bak,$^{ \negthinspace  a}$  Chanju Kim,$^{ \negthinspace b}$ Sang-Heon Yi,$^{\negthinspace c}$} }
\vspace{0.8cm} 
\centerline{\sl  a) Physics Department \& Natural Science Research Institute}
\centerline{\sl University of Seoul, Seoul 02504 \rm KOREA}
 \vskip0.2cm
 \centerline{\sl b) Department of Physics, Ewha Womans University,
  Seoul 03760 \rm KOREA}
   \vskip0.2cm
 \centerline{\sl c) Center for Quantum Spacetime \&  Physics Department}
  \centerline{\sl Sogang University,  Seoul 04107 \rm KOREA}
\vskip0.4cm

 \centerline{
\tt{(\small dsbak@uos.ac.kr,\,cjkim@ewha.ac.kr,\,shyi@sogang.ac.kr})
} 
  \vspace{1.5cm}
%\centerline{\today}
%\vspace{1.75cm}
\centerline{ABSTRACT} \vspace{0.65cm} 
{
\noindent 
We consider Jackiw-Teitelboim gravity with a massless matter field  and turn 
on bulk excitations leading to a nontrivial vev of the corresponding dual 
boundary operator. To leading order, we realize the corresponding 
deformation of thermofield double state by explicitly identifying their 
Hilbert space. The deformed state can be prepared with an operator insertion
at the mid-point of the Euclidean time evolution in the context of
Hartle-Hawking construction. We show that the inserted operators 
form an SL(2,{\bf R}) representation. We construct a specific orthonormal
basis that is directly related to the operator basis of the vev deformations.
If we include the higher order corrections, the bulk geometry is no longer
left-right symmetric. We argue that, classically, the mode coefficients in the bulk
deformation cannot be fully recovered from the data collected along the  boundary cutoff
  trajectories. %It implies that 
Then the bulk seems to contain more information
than the cutoff  boundary, and this might be responsible for nontrivial
behind-horizon degrees of freedom.

%We find general deformations of BTZ spacetime and identify the corresponding thermofield initial
%states of the dual CFT. We deform the geometry by introducing bulk fields dual to primary
%operators and find the back-reacted gravity solutions to the quadratic order of the deformation
%parameter. The dual thermofield initial states can be deformed by inserting arbitrary linear
%combination of operators at the mid-point of the Euclidean time evolution that appears in the
%construction of the thermofield initial states. The deformed geometries are dual to thermofield
%states without deforming the boundary Hamiltonians in the CFT side. We explicitly demonstrate
%that the AdS/CFT correspondence is not a linear correspondence in the sense that the linear
%structure of Hilbert space of the underlying CFT is realized nonlinearly in the gravity side.
%We also find that their Penrose diagrams are no longer a square but elongated horizontally
%due to deformation. These geometries describe a relaxation of generic initial perturbation of
%thermal system while fixing the total energy of the system. The coarse-grained entropy grows
%and the relaxation time scale is of order β/2π. We clarify that the gravity description involves
%coarse-graining inevitably missing some information of nonperturbative degrees.

}

%\vspace{0.75cm}
%\centerline{(\today)}
\end{titlepage}
%%%%%%%%%%%%%%%%%%%%%%
%\maketitle

%%%%%%%%%%%%%%%%%

%%%%%%%%%%%%%%%%%%%%%%%%%%%%%%%%%%%%%%%%%%%%%%%%%%%
\section{Introduction
}\label{sec1}
%%%%%%%%%%%%%%%%%

In recent years, there have been remarkable developments  in the context of the Nearly AdS$_{2}$/Nearly CFT$_{1}$\,(NAdS$_{2}$/NCFT$_{1}$) correspondence. NAdS$_{2}$ %geometry  
arises
from an appropriate dimensional reduction of %higher-dimensional 
a near extremal black hole geometry while 
 NCFT$_{1}$ may appear as a low energy approximation of a one-dimensional quantum system like the SYK model (See~\cite{Sarosi:2017ykf} for a review and also  references therein.). 
%emerges as a background geometry for near horizon dynamics of some extremal black holes through dimensional reduction, and NCFT$_{1}$ may appear as a low energy approximation of a one-dimensional quantum system like the SYK model (See~\cite{Sarosi:2017ykf} for a reivew).  
%This correspondence spawned  many interesting (cooperative) works between high energy  physics and condensed matter physics communities. 
%The Jackiw-Teitelboim model coupled with matters provides a concrete Lagrangian description of NAdS$_{2}$ dynamics and its boundary dynamics is reduced to the Schwarzian one. 
The Jackiw-Teitelboim (JT) model (coupled with a matter field) \cite{Jackiw:1984je,Teitelboim:1983ux,Almheiri:2014cka}
is a specific 2d dilaton gravity for the NAdS$_2$
geometry, which  %whose dynamics along boundary cutoff trajectories %, in particular, 
may be reduced to the  Schwarzian  dynamics along the boundary cutoff trajectories 
\cite{Maldacena:2016upp}.
%Furthermore, 
Combined with  various %(quantum) 
information-theoretic techniques, this model provides a computable test bed for various ideas about the  resolution of the black hole information loss problem~\cite{Almheiri:2020cfm}.

%In the finite temperature context,  a  static AdS black hole solution is known to be well-matched with  the so-called thermofield double state as its dual. And the normalizable and non-normalizable modes of bulk matters correspond to  the vev  and source  of their boundary dual operators  respectively, in the standard quantization\footnote{In a certain range of mass parameter of a bulk field, there could be an alternative quantization which is not our concern here~\cite{Breitenlohner:1982jf,Witten:2001ua}.}. Hence, the normalizable mode of a small bulk matter deformation on a black hole corresponds to the state deformation of its dual thermofield double state. Moreover, this state deformation can be mapped to an operator insertion on the boundary of Euclidean AdS$_{2}$ by the state-operator mapping, where the Euclidean AdS$_{2}$ is introduced by the Hartle-Hawking construction for the preparation of the thermofield double state.  

A two-sided AdS black hole geometry is well known to be dual to the so-called 
thermofield double state 
in the boundary side \cite{Maldacena:2001kr}.  The (undeformed) thermofield initial state may be prepared by the  Euclidean time  evolution 
 in the context of
Hartle-Hawking construction %.
%along
%the boundary of %half 
% Euclidean AdS$_2$ 
\cite{Maldacena:2001kr}.
In this note, we are mainly interested in  rather general deformations of the two-sided black hole geometry by turning on a bulk matter field that is dual to the corresponding boundary operator.
%There are two independent
 %This undeformed thermofield initial state may be prepared by a  Euclidean evolution along
%the boundary of (a half of) Euclidean AdS$_2$
%The normalizable 
% and non-normalizable modes of the bulk scalar field correspond to  the vev  and source  of their boundary dual operators  respectively, in the so-called standard quantization\footnote{In a certain range of mass parameter of a bulk field, there could be an alternative quantization which is not our concern in this note~\cite{Breitenlohner:1982jf, Witten:2001ua}.}. 
In the so-called standard quantization\footnote{In a certain range of mass parameter of a bulk field, there could be an alternative quantization which is not our concern in this note~\cite{Breitenlohner:1982jf, Witten:2001ua}.}\negthinspace\negthinspace, 
the normalizable modes of the bulk matter field correspond to  the vev's  of the boundary operator
 while the non-normalizable modes are dual to %of the bulk scalar field correspond to  the vev  and 
the source  deformations of the boundary theory by the same boundary  operator.
 % respectively, 
Thus any excitation  of a black hole geometry by the normalizable modes %of a small bulk matter 
%deformation 
will lead %to %corresponds 
to the corresponding 
 deformations of the thermofield double state. % in an appropriate manner.
In this note, specialized to the case of the massless scalar field, we would like to clarify the structure of deformations in the 2d bulk %geometries
as well as in the dual boundary theory side. Especially, we shall show that the initial state, to the leading order of vev deformations, may be prepared with an operator insertion
at the mid-point of the Euclidean time evolution \cite{Goto:2017olq, Bak:2017xla}.
% in the context of
%Hartle-Hawking construction.
 %Moreover, this state deformation can be mapped to an operator insertion on the boundary of Euclidean AdS$_{2}$ by the state-operator mapping, where the Euclidean AdS$_{2}$ is introduced by the Hartle-Hawking construction for the preparation of the thermofield double state.  

In general, deformations would affect the dynamics of the cutoff trajectories.
One may then try to obtain the information of the deformations
by probing the cutoff trajectories at the boundary.
This would, in principle, be possible if the trajectories contain
all the information of the bulk deformations. There are, however,
nontrivial behind-horizon degrees of freedom in the bulk such as Python's
degrees of freedom \cite{Brown:2019rox,Engelhardt:2021mue,Bak:2021qbo}. It is not clear at all whether this hidden information
can be fully recovered by collecting boundary data. In fact, by explicitly
solving the equations of motion, we will see that the cutoff trajectories
do not have enough information for the full recovery.

Since AdS$_{2}$ is rigid even under the bulk matter deformation, the SL$(2,{\bf R})$ symmetries of the background geometry provide useful information for understanding the relevant dynamics~\cite{Lin:2019qwu}. 
In this paper, we explore the SL$(2,{\bf R})$ symmetry  realization of the inserted operators corresponding to the vev  deformations of the thermofield double state. To simplify the discussion, we consider  %restrict ourselves to 
these %small 
vev
deformations only up to %in 
their leading order. % ignoring higher order corrections. %to the thermofield double state. 
We show %that the operator insertion resides at the mid-point of the Euclidean time evolution and 
that these inserted operators form a specific SL$(2,{\bf R})$ representation. See~\cite{Kitaev:2018wpr, Lin:2019qwu,Lin:2022rbf} 
for some related discussions.  

This paper is organized as follows. In the next section, we review the Jackiw-Teitelboim model focusing on the 2d two-sided black hole geometries and the induced Schwarzian dynamics 
along their cutoff trajectories. In Section 3, we 
 investigate the structure of generic bulk deformations by turning
on a massless scalar field. We argue that the bulk information may not be fully
recovered from the boundary data collected along the cutoff trajectories.
Sections 4 and 5 are devoted to the 
realization of SL$(2,{\bf R})$ symmetries of the inserted operators. 
To the leading order, we explicitly identify the operators inserted at the 
mid-point which reproduce the most general vev deformations and show that 
they form a unitary SL(2,{\bf R}) representation.
In the final section, we summarize our results and comment on  some future directions.

%%%%%%%%%%%%%%%%%
\section{Two-dimensional dilaton gravity 
}\label{sec2}
%%%%%%%%%%%%%%%%%

%Jackiw-Teitelboim 
The JT
model of our interest
is a 2d dilaton gravity with a matter field described by action
\bea
I=I_{top}+\frac{1}{16\pi G}\int_M d^2 x \sqrt{-g}\, \phi \left( R+\frac{2}{\ell^2}\right) + I_{surf}  + I_M(g, \chi) \,,
\label{euclidaction}
\eea
where $\phi$ is a dilaton field, $\chi$ a matter field and
\begin{align}    \label{}
I_{top}&=  \frac{\phi_0}{ 16\pi G}\int_M d^2 x \sqrt{-g}  R\,,  \nonumber \\
I_{surf} &=   \frac{1}{ 8\pi G}\int_{\partial M} \sqrt{\gamma}\, (\phi_0 + \phi ) \, K \,, \nonumber \\
I_M &= -\frac{1}{2}\int_M d^2 x \sqrt{-g} \left( \nabla \chi \cdot \nabla \chi + m^2 \chi^2 \right)\,.
\end{align}
In this action, $\ell$ is the AdS radius, and $\gamma_{ij}$ and $K$
denote the induced metric and  the extrinsic curvature on the boundary 
$\partial M$, respectively.

The variation of the dilaton field $\phi$ leads to
\begin{equation}
R+\frac{2}{\ell^2}=0\,,
\end{equation}
which sets the metric to be AdS$_2$. The other equations of motion
are obtained from the variation of the metric $g$ and the scalar field $\chi$,
\begin{align}    \label{phieq}
\nabla_a \nabla_b \phi -g_{ab} \nabla^2 \phi + \frac{1}{\ell^{2}}g_{ab} \phi &= - 8 \pi G T_{ab}\,, \\
\nabla^2 \chi -m^2 \chi &=0\,,  \label{chieq}
\end{align}
where
$T_{ab}$ is the stress tensor of the matter field,
\begin{equation} \label{}
T_{ab} = \nabla_a \chi \nabla_b \chi  -\frac{1}{2} g_{ab} \left( \nabla \chi \cdot \nabla \chi + m^2 \chi^2 \right)\,.
\end{equation}
In the global coordinates, the metric of the AdS$_2$ space is written as
\begin{equation} \label{}
ds^2 =\frac{\ell^2}{\cos^2 \mu} \left(-d\tau^2 + d\mu^2  \right)\,,
\end{equation}
where $\mu \in [-\frac{\pi}{2},\frac{\pi}{2}]$. The most general
vacuum solution to the dilaton equation of motion 
\eqref{phieq} with $T_{ab}=0$ is given by
\begin{equation}
\phi= \phi_{BH}(L,b,\tau_{B})\equiv \bar\phi \, L\,\,\frac{(b+b^{-1}) \cos (\tau-\tau_B) -(b-b^{-1}) \sin \mu}{2 \cos \mu} \,,
\label{dilaton}
\end{equation}
By the coordinate transformation 
\begin{align}    \label{}
\frac{r}{L} &= \frac{(b+b^{-1}) \cos (\tau-\tau_B) -(b-b^{-1}) \sin \mu}{2\cos \mu}\,,  \nonumber \\
 \tanh \frac{t L }{\ell^2} &=\frac{2\sin (\tau-\tau_B)}{(b+b^{-1}) \sin \mu -(b-b^{-1}) \cos (\tau-\tau_B)}\,,
 \label{coorb}
\end{align}
we obtain the AdS black hole metric
\begin{equation} \label{btz}
ds^2= - \frac{r^2-L^2}{\ell^2} dt^2+ \frac{\ell^2}{r^2-L^2} dr^2\,,
\end{equation}
with $\phi = \bar\phi \, r$.
%
%The  Penrose diagram for the above black hole with $b=1$ is depicted  in Figure \ref{fig02}. % in ($\mu,\tau$) space. 
Utilizing the SL(2,\,{\bf R}) isometry of AdS$_2$, we can set
$b=1$ and $\tau_B=0$ \cite{Bak:2018txn}.
This metric describes the Rindler wedge of two-sided AdS black holes with
the radius of black hole horizon $L$. 
The location of singularity is defined by the curve 
$\Phi^2\equiv \phi_0 +\phi =0$ in the above dilaton field, and  $\Phi^2$ 
might be viewed as characterizing the size of  ``transverse space"~\cite{Almheiri:2014cka}.  
%
%{\color{blue}
%$b=0$ is the Poincare AdS and work out the coordinate transformation form global AdS to Poincare AdS.
%Draw also 
%the shape of singularity that is defined by $\Phi^2 \equiv \phi_0 +\phi =0$ for each case.
%}
%\section{Thermodynamics of AdS black hole}
%
In this left/right symmetric two-sided black hole case, one can see that
the Gibbons-Hawking temperature, the entropy  and energy  are given by 
\begin{equation} \label{}
T= \frac{1}{2\pi} \frac{L}{\ell^2}\,,  \qquad S= S_0 +{\cal C} T\,,   \qquad 
E = \frac{1}{2} {\cal C}T^2\,,
\end{equation}
where $S_0$ is the ground state entropy given by $S_{0}= \frac{\phi_0} {4G}$ and ${\cal C} =\frac{\pi \bar\phi \ell^2}{2 G}$. In general, these physical quantities could be different for left/right Rindler wedges  in the two-sided black hole case. In the next sections  we  consider some deformation of black hole configuration through the dilaton field and show that  these quantities are indeed different  for %from 
the left/right Rindler wedges.

The boundary time $u$ may be introduced 
in $\epsilon \rightarrow 0$ limit through the prescription
\begin{equation} \label{dscutoff}
ds^2|_{\text{cutoff}} = -\frac1{\epsilon^2} du^2, \qquad
\phi|_{\text{cutoff}} = \bar\phi \frac\ell\epsilon.
\end{equation}
We will adopt the convention that the right boundary time $t_r$ runs upwards
whereas the left boundary time $t_l$ runs downward. In other words, we identify
\begin{equation} \label{bdryu}
u=t_r = -t_l.
\end{equation}
See Section \ref{sec4} for the details.

The boundary dynamics is equivalently described by a Schwarzian 
theory~\cite{Jensen:2016pah, Engelsoy:2016xyb, Maldacena:2017axo},
\begin{equation} \label{schwarzian}
S=\int du \left[ -\phi_l \left\{ \tan \frac{\tau_l(u)}2,u \right\}
                -\phi_r \left\{ \tan \frac{\tau_r (u)}2,u \right\} \right],
\end{equation}
where $\phi_l=\phi_r$ can be identified with $\bar\phi$ in the bulk and $\tau_{l/r}(u)$ corresponds to the left/right global time coordinate, respectively,  at each cutoff trajectory.
If the matter is turned on, 
%the relation \eqref{bdryu} and 
the Schwarzian action would get corrections which, in general, 
can be asymmetric at the left and the right boundaries.
To the leading order in the deformation, however, the correction 
vanishes, as seen in the next section.
%so that \eqref{bdryu} holds, 

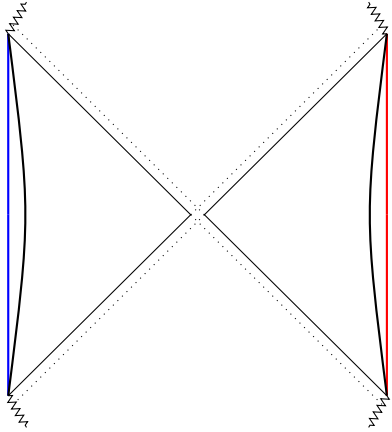
\begin{figure}[htbp]  
\vskip0.3cm 
\begin{center}
\begin{tikzpicture}[scale=1.2]
\draw[thick,blue %,-<
](-2,0)--++(0,2);
\draw[thick,blue] (-2,2)%node[left]{\large$t_l$}
--++(0,2);
\draw[thick,red %, ->
](2.15,0)--++(0,2);\draw[thick,red] (2.15,2)%node[right]%{\large$t_r$}
--++(0,2);
%\draw[blue,decoration={coil,segment length=0.5mm,amplitude=0.15mm},decorate] (-2,2) arc (180:270:2);\draw[red,decoration={coil,segment length=0.5mm,amplitude=0.15mm},decorate] (2,2) arc (0:-90:2); \draw[fill=black] (0,0) circle (0.05cm);
%\draw[decoration={zigzag},decorate](-2,0)--(2,0); 
% \draw[thick,dotted](-2,2)--++(4,0);
%\draw[decoration={zigzag},decorate](-2,4)--++(4,0); 
\draw[%violet,
decoration={zigzag,segment length=1mm,amplitude=0.5mm},decorate
](-2,4)--++(0.2,+0.35); 
\draw[decoration={zigzag,segment length=1mm,amplitude=0.5mm},decorate
](2.15,4)--++(-0.2,+0.35); 
\draw[decoration={zigzag,segment length=1mm,amplitude=0.5mm},decorate
](-2,0)--++(0.2,-0.35); 
\draw[decoration={zigzag,segment length=1mm,amplitude=0.5mm},decorate
](2.15,0)--++(-0.2,-0.35); 
%\draw[thick,brown,dotted] (-2,0)--++(4,4);
%\draw[thick,brown,dotted] (+2,0)--++(-4,4);
\draw (-2,0)--++(2,2)--++(-2,2);
\draw[dotted] (0,2)--++(2.08,2.08);\draw[dotted] (0,2)--++(2.10,-2.10);
\draw[dotted] (0.15,2)--++(-2.10,2.10);\draw[dotted] (0.15,2)--++(-2.08,-2.08);
\draw (+2.15,0)--++(-2,2)--++(2,2);
%\draw (+2,0)--++(-2,2)--++(2,2);
%\draw[blue,decoration={coil,segment length=0.5mm,amplitude=0.15mm},decorate] (6,4) arc (90:270:2);%\draw[red,decoration={coil,segment length=0.5mm,amplitude=0.15mm},decorate] (6,4) arc (90:-90:2); %\draw[fill=black] (6,0) circle (0.05cm);
 %\draw[thick,dotted](4,2)--++(4,0);\draw[fill=black] (6,0) circle (0.05cm);
%\draw[decoration={zigzag},decorate](-2,4)--++(4,0); 
%\draw[blue,decoration={coil,segment length=0.5mm,amplitude=0.2mm},decorate] (4,2) arc (180:90:2);
%\draw[red,decoration={coil,segment length=0.5mm,amplitude=0.2mm},decorate] (8,2) arc (0:90:2); 
%\draw[fill=black] (6,4) circle (0.05cm);
%\tikz[every to/.style={bend right}]
% \draw[every to/.style={bend right}] (-2,0) to (-2,4);
\draw[thick] (-2,0) .. controls (-1.75,2) .. (-2,4);
% \draw[every to/.style={bend left}] (2,0) to (2,4);
\draw[thick] (2.15,0) .. controls (1.9,2) .. (2.15,4);
\end{tikzpicture}
\end{center}
\vskip-0.3cm 
\caption{\small The left and the right cutoff trajectories are illustrated 
as curves near the boundaries in the Penrose diagram of a typical 
deformed space.}\label{Janus}
\end{figure}

We depict the Penrose diagram of a deformed space in Figure~\ref{Janus}.
In the figure, the curves near the boundaries represent typical
cutoff trajectories of the boundary dynamics. Given dilaton 
configurations that are deformed away from \eqref{dilaton}, 
one can obtain the cutoff trajectories by using
the prescription \eqref{dscutoff} or from the boundary action with
deformed terms.

%%%%%%%%%%%%%%%%%%%%%%%%%%%%%%%%%%%%%%%%%%%%
%\subsection*{Deformations}\label{sec2.1}
%%%%%%%%%%%%%%%%%%%%%%%%%%%%%%%%%%%%%%%%%%%

% %
%
%
%In this section we shall construct various full back-reacted solutions describing thermalization of initial perturbations
%of black holes and investigate their general properties. It will be later on used to understand  the dynamics of teleportee through the bulk.  We will set $\ell=1$ in this and the next section. 
%The corresponding Penrose diagram will be a rectangle instead of square. It will be a rectangle
%elongated horizontally in general.

Now we briefly review the general deformation by the matter 
field. As a simple left-right asymmetric case,
let us recall the eternal Janus deformation~\cite{Bak:2018txn} 
which makes the Hamiltonians of left-right boundaries differ from each 
other by turning on exact marginal operators. It is given by
\begin{equation} \label{ejanus}
\chi =\gamma (\mu-\kappa_{0})\,, \qquad \phi = \bar{\phi}L \frac{\cos\tau}{\cos\mu} -4\pi G\gamma^{2}(1+\mu\tan\mu)\,.
\end{equation}
%
%where $-/+$ signature for the left/right side, respectively. 
In this case, 
though the matter field $\chi$ is asymmetric, the dilaton field $\phi$ and 
the black hole temperature are left-right symmetric under the exchange of 
$\mu \leftrightarrow -\mu$. 

Since the metric is fixed to be AdS$_2$, the matter field equation 
\eqref{chieq} can be solved. In the global coordinates, the general solution 
is given by~\cite{Spradlin:1999bn}
\begin{equation} \label{gdeltasol}
\chi= \sum^\infty_{n=0} c^{\cal D} _n \, {\cal N}_n \cos^{\cal D} \mu \, C_n^{\cal D} (\sin \mu) \, e^{-i(n+{\cal D})\tau %+\tau_{0})
}  + {\rm c.c.}\ ,
\end{equation}
where 
\begin{equation} \label{}
{\cal N}_n =2^{{\cal D} -1} \Gamma({\cal D}) {\textstyle  \sqrt{\frac{\Gamma(n+1)}{\pi \Gamma(n+2{\cal D})}}}\,,
\end{equation}
and $C^{\cal D}_n(x)$ denotes the Gegenbauer polynomial. %defined in \cite{grad}. 
Here, the parameter ${\cal D}$ is defined in terms of the mass of the 
scalar field $\chi$ as
\begin{equation} \label{calD}
{\cal D}=\Delta_\pm= \frac{1}{2} \left( 1 \pm \sqrt{1+4m^2}\right)\,.
\end{equation}
According to the AdS/CFT correspondence, the bulk matter $\chi$ is dual to
a scalar primary operator $O_\Delta(t)$ of a certain dimension $\Delta$.
If $m^2 \ge 0$, it is identified as       $\Delta_+$, {\it i.e.}, 
\begin{equation} \label{}
{     \Delta = \Delta_{+} }=\frac{1}{2} \left( 1 + \sqrt{1+4m^2}\right)\,.
\end{equation}
In this case, the deformation by the matter $\chi$ with 
 $     {\cal D}=\Delta_{+} $ in    \eqref{gdeltasol}  corresponds to a vev deformation 
of the dual field theory, while the other deformation with identifying   
${\cal D}=\Delta_{-}$  describes a source deformation. 
In AdS$_2$, there is another possibility that $m^2$ is negative as long as the 
Breitenlohner-Freedman bound~\cite{Breitenlohner:1982jf} holds, namely
$ -1/4 \le m^2  < 0 $ in our case. Then, we have two possible values
as the operator dimension $\Delta$,
\begin{equation} \label{}
\Delta =\Delta_\pm= \frac{1}{2} \left( 1 \pm \sqrt{1+4m^2}\right)\,.
\label{deltapm}
\end{equation}

\section{Structure of bulk deformations %with a massless bulk field
%and multiple bulge geometries %($m^2=0$) 
}\label{sec3}
From this section, we shall consider turning on a massless scalar field %in the bulk 
dual to a scalar operator of dimension $\Delta=1$. In this case, the bulk scalar equation  (\ref{phieq}) can be solved 
explicitly by \cite{Bak:2021qbo}
\bea\label{fullscalar} 
\chi = \chi_s +\chi_v\,,
\eea
where
\be \label{chiv}
\chi_v=\sum^\infty_{n=1}a_{n}\sin n\big(\mu+{\frac{\pi}{2}}\big) \cos n(\tau-\tau^v_n) 
%+\sum^\infty_{n=0}b_{2n+2}\sin (2n+2)\mu \cos (2n+2)\tau 
\ee 
and
\be \label{chiv}
\chi_s=b_0 +\sum^\infty_{n=1}b_{n}\cos n\big(\mu+{\frac{\pi}{2}}\big) \cos n(\tau-\tau^s_n) \,.
%+\sum^\infty_{n=0}b_{2n+2}\sin (2n+2)\mu \cos (2n+2)\tau 
\ee 
This describes a fully general set of classical solutions to the scalar equation in the ambient 
AdS$_2$ space where $\chi_{v/s}$ satisfies %respectively 
the Dirichlet/Neumann boundary condition at 
$\mu=\pm \pi/2$. In our JT model, the metric remains always to be  AdS$_2$ and won't be corrected by any matter perturbations. For the dilaton field, we shall start from the vacuum 
solution
\be %\label{}
\phi_{bg} =\, \bar\phi L \,
\frac{\cos \tau}{\cos \mu} 
\ee
without loss of any generality. As was mentioned previously,  this background 
describes a two-sided black hole with temperature $T=\frac{1}{2\pi}\negthinspace\frac{L}{\,\,\ell^2}$.
With the scalar deformation (\ref{fullscalar}), the dilaton solution becomes
\be \label{fullsoln}
\phi = \phi_{bg} + %\varphi
%\ee
%with
%\be
%\varphi=
 %+8\pi G \sum^\infty_{n=1}b_{2n}^2 \phi_{2n}+
8\pi G  \sum^\infty_{m=1}\sum^\infty_{n=1}\left(\, a_{m}a_{n} \phi^v_{m,n} +b_{m}b_{n} \phi^s_{m,n} +
2a_{m}b_{n} \phi^c_{m,n} \, \right)\,,
\ee
%with
%\begin{align}
%\phi_{cross} =
%\end{align}
where the full detailed functional forms of $\phi^{v,s,c}_{\, m,n}$ are given explicitly in \cite{Bak:2021qbo}.

We now turn to the standard AdS/CFT interpretation of the above bulk deformation. To this end, note that 
the scalar solution in any asymptotic region of  Lorentzian AdS$_2$ spacetime may be expanded as\footnote{ In this expansion, we have ignored any logarithmic terms which are not relevant in our discussion below.}  (See \cite{Skenderis:2002wp} for instance)
\be
\chi_\Delta =s(\tau) \left(\frac{\ell^2}{r}\right)^{1-\Delta} (1+ \cdots)
+u(\tau) \left(\frac{\ell^2}{r}\right)^\Delta (1+ \cdots)~,
\ee
where the radial coordinate $r$ is defined by $\phi/\bar\phi$ and $\cdots$ denotes higher order contributions  of each power series expansion in
$\left(\ell^2 /r\right)^2$. 
In this note, we are dealing with deformations of the two-sided black hole which involves
%the 
two (left and right) asymptotic regions in general. 
In the left/right asymptotic region, the presence of %the source term
nonvanishing 
$s_{l/r}(\tau_{l/r}(u))$ represents a source deformation of the left/right boundary theory by
\be \label{lagrangian}
{\cal L}_{l/r}(u) = {\cal L}_0 (u)+ s_{l/r}(\tau_{l/r}(u)) O_\Delta (u)\,,
\ee
where $\tau_{l/r}(u)$ describes the reparameterization along the left/right cutoff trajectory respectively. Below for the simplicity of our presentation, we shall write $\tau_{l/r}(u)$ simply as $\tau$ %frequently 
once it is not confusing.
With our normalization in (\ref{euclidaction}), the vev of operator with $s=0$ may be identified
as \cite{Skenderis:2002wp}
\be
\langle O^{l/r}_{\Delta}(u)\rangle_{s=0}=%\frac{1}{2}  
(2\Delta -1) u_{l/r}(\tau)\,. %(\tau_{l/r}(u))
\ee
In our case of $\Delta=1$, we shall also introduce normalized vev functions $w_{l,r}(u)$ by
\be
\label{vevdef}
\langle O_{l/r}(u) \rangle_{s=0}=u_{l/r}(\tau) \equiv \frac{2\pi}{\beta} w_{l/r}(u)\,.
\ee
From the above solution in (\ref{fullscalar}), the source term for our $\Delta =1$ case may be identified as
\be\label{sourcen}
s_{l/r}=b_0 +\sum^\infty_{n=1}(\pm 1)^n b_{n} \cos n(\tau%(\tau_{l/r}(u))%\tau_{l/r}(u)
-\tau^s_n) \,.
%+\sum^\infty_{n=0}b_{2n+2}\sin (2n+2)\mu \cos (2n+2)\tau 
\ee 
Here and below, the upper/lower sign is  for the left/right system respectively. Similarly the vev function can also be identified as
\be
w_{l/r} =Q_{l/r}(\tau)\sum^\infty_{n=1}(\pm 1)^{n+1} n \, a_{n} \cos n(\tau-\tau^v_n) \,,
\ee
where $Q_{\l/r}$ is defined by
\be
Q_{l/r} (\tau) =\lim_{\mu \rightarrow \mp \pi/2} \frac{\phi \cos \mu}{ \bar\phi \, L}\,.
\ee
Without the source deformation, the $Q_{l,r}$ functions may be computed as \cite{Bak:2021qbo}
\be
Q_{l/r} (\tau) =\sqrt{A^2_{l/r}+B^2_{l/r}} \,\,\Bigl(\, \cos(\tau-\tau_{l/r}^B) -q_{l/r}\Bigr)\,,
\ee
where
\bea
A_{l/r}=1+{\cal O}(a^2), \ \  B_{l/r}={\cal O}(a^2), \ \  q_{l/r}={\cal O}(a^2),\ \ \tan \tau_{l/r}^B = \frac{B_{l/r}}{A_{l/r}}={\cal O}(a^2)\,,
\eea
whose precise functional forms are also given in  \cite{Bak:2021qbo}.
Once this function  $Q_{l/r}$ is given, the left/right reparameterization dynamics may be  solved  explicitly by \cite{Bak:2021qbo}
\be
\cos(\tau-\tau_{l/r}^B) -q_{l/r} =\frac{1-q^2_{l/r}}{ \cosh \frac{2\pi u}{\phantom{ii}\beta_{l/r}}   +q_{l/r}}
\ee
with the temperature
\be
T_{l/r} =T \sqrt{A^2_{l/r}+B^2_{l/r}}\sqrt{1-q^2_{l/r}}=T \, \big(1+{\cal O}(a^2)\big)\,.
\ee
One is then led to
\be 
Q_{l/r}=\frac{\sqrt{1-q^2_{l/r}}}{ \cosh \frac{2\pi u}{\phantom{ii}\beta_{l/r}}   +q_{l/r}} \frac{{\beta} }{{\phantom{1}\beta_{l/r}} }\,.
\ee
Thus from these overall factors in the vevs, one may see that the vevs decay exponentially in the boundary time
$u$,
whose decay rate  is precisely  the expected one with our dimension one operator. 
With these vev deformations, the boundary Schwarzian dynamics are basically those of asymptotically 
black hole spacetimes. 

By turning on source deformations only, the boundary reparameterization dynamics may be solved in a similar manner. The resulting 2d spacetime describes again left-right asymmetric black holes in general. If one turns on both the source and  the vev deformations at the same time, the left  and  right black holes are further excited, which will be reflected in the boundary Schwarzian dynamics by the excitation of the corresponding reparameterization modes. In any of the deformations mentioned in the above, one may show that 
$\tau_{l/r}(u=\infty)-\tau_{r/l}(u=-\infty) \le \pi$, which implies that one cannot send a signal from one side to the other~\cite{Bak:2021qbo}. Namely the two boundaries are causally disconnected from each other.
From the view point of boundary systems, the left and   right systems are completely decoupled from each other
without any direct interactions  to permit any information transfer  between them. 
Note further  that in general the left and right black holes become different from each other as a result of deformation. Especially their temperatures become different from each other in general. 

From now on, let us focus on the above deformations only to the leading order,  ignoring any higher order corrections. In this limit, the left-right black holes remain unperturbed with $T_{l}=T_r=T$. On the left/right cutoff trajectory, $\tau_{l/r}(u)$ is ranged over $(-\pi/2, \pi/2)$ and the reparameterization is solved by $\sin \tau_{l/r} = \tanh \frac{2\pi }{\beta}u$, respectively. 
The left and right source terms $s_{l,r}$ take the same forms as (\ref{sourcen}) and the vev functions $w_{l,r}$ become
\be
\label{leadingvev}
w_{l/r} =\cos\tau \,\sum^\infty_{n=1}(\pm 1)^{n+1} n \, a_{n} \cos n(\tau-\tau^v_n) \,.
\ee
Let us first show the independence of the left  and right perturbations. By defining
\be
s(\tau) =\left[ \begin{array}{lll}
s_r(\tau)   & \ \ \ {\rm for} & -\frac{\pi}{2} < \tau < \frac{\pi}{2} \\
  s_l(\tau-\pi)  & \ \ \   {\rm for}      &  \phantom{ii}\frac{\pi}{2} < \tau < \frac{3\pi}{2}
 \end{array} \right.
\ee
one then finds
\be
s(\tau)=b_0 +\sum^\infty_{n=1}(-1)^n b_{n} \cos n(\tau%(\tau_{l/r}(u))%\tau_{l/r}(u)
-\tau^s_n) 
\ee
which describes a general $2\pi$-periodic real function defined over the range $(-\pi/2, 3\pi/2)$. This implies that the source functions $s_{l,r}$ defined over the range $(-\pi/2, \pi/2)$ become totally independent from each other. This of course agrees with the fact that the source terms in the left  and right boundary actions can be turned on independently from each other.
By a similar argument, one may show that the vev functions $w_{l,r}$ may also be turned on %totally 
independently from each other. 

One straightforward consequence of the above consideration is that one may in principle recover the full set of source mode-coefficients
$\{ b_n e^{-i n\tau^s_n} \}$ from the left  and right boundary data specified by $s_{l,r}(\tau)$
with $\tau \in (-\pi/2,\pi/2)$.  (Of course a similar statement can also be made for the vev deformations.) Therefore, in the small deformation limit (i.e. working in the leading order of the above deformations), probing the left  and right boundary perturbations all together, one may recover the corresponding  mode-coefficients completely.  However, including the higher order correction in bulk geometries,   the situation changes completely. 
Due to the shifts in $\tau_{l/r}(\pm\infty)$, the total interval size $\Delta \tau_l + \Delta \tau_r$   of the left right cutoff trajectories in $\tau$ space becomes in general less than $2\pi$ %in general
 where $\Delta \tau_{l/r}$ denotes $\tau_{l/r}(\infty)-\tau_{l/r}(-\infty)$, respectively. 
This implies that
the mode-coefficients cannot be fully recovered from the left and right boundary 
data collected along the full trajectories. 
On the other hand, one may try to investigate the corresponding %$\mu$-dependent 
bulk profiles of the scalar field.
%from the bulk $\mu$ dependence of $\chi(\mu,\tau)$
For instance  consider the two-independent bulk functions $\chi_{s/v} (\mu,\tau_1)$ and $\chi_{s/v} (\mu,\tau_2)$ with an appropriate choice of $\tau_1$ and $\tau_2$ ($\tau_1 \neq \tau_2$), for which the Neumann/Dirichlet boundary condition is imposed at $\mu=\pm\pi/2$ for the source/vev deformation, respectively.  From these two independent functions, one may verify that the mode-coefficients  can be recovered
 completely\footnote{This does not imply an existence of  a bulk observer who may collect all the required information for the recovery while traveling in  the bulk.}.  Thus it seems that the bulk in this case contains more information than the ones that may be probed from the boundaries. One may speculate that this hidden information in the bulk is responsible for those nontrivial behind-horizon degrees of freedom such as Python's lunch degrees of freedom discussed in \cite{Bak:2021qbo}.

In the next section, we shall identify the deformed boundary states from which the above 
leading-order bulk gravity results follow precisely.

%%%%%%%%%%%%%%%%%%%%%%%%%%%%%%%%%%%%%%%%%%%%%%
\section{Deformation of thermofield double state %($m^2=0$) 
}\label{sec4}
%%%%%%%%%%%%%%%%%%%%%%%%%%%%%%%%%%%%%%%%%%%%%
In this section, we shall introduce the thermofield double state~\cite{Takahasi:1974zn} in the boundary theory and its deformations that reproduce the above mentioned bulk results to the leading order. Especially, we would like to focus on the %detailed 
state %structure of 
deformations which lead to the vev functions in (\ref{vevdef}) and (\ref{leadingvev}). Below we shall also discuss their relation to the bulk gravity description. % and their information theoretic aspects.
 In these vev deformations, we expect that the SL(2,\bR) symmetries of the boundary system will be realized 
with a certain Hilbert space representation~\cite{Kitaev:2018wpr,Lin:2019qwu}. 
In the next section we shall present a specific form of orthonormal basis that is directly related to the operator basis of the vev deformations. 

We shall begin with the thermofield double initial state in the boundary theory defined by
 \begin{equation} \label{TFD1}
|\Psi\rangle = \frac{1}{\sqrt{Z}}\sum_{m,n} \langle n| \, U \,|m\rangle~ |\bar{m}\rangle_{l} |n\rangle_{r}\,,  \qquad  Z={\rm tr}\, U^\dagger U\,,
\end{equation}
where $U$ is a Euclidean evolution operator that will be further specified  below and $|\bar{m}\rangle$  denotes
the CTP conjugated state of a basis state $|m\rangle$. In our convention, any operators 
labeled by $l/r$ will act on the left/right Hilbert space respectively with an extra transpose operation %$T$ 
in the case of left side operators.
The Lorentz time evolution of this initial state  $|\Psi\rangle$ 
is given by
\begin{equation} \label{lorentzianevol}
|\Psi(t_{l},\, t_{r})\rangle = %{\cal T}_l\,\text{exp} \Big[\, i\int^{t_{l}}_{0} dt_{l}\,  H_l(-t_{l})\Big]
{\cal T}_l\,e^{ i\int^{t_{l}}_{0} dt_{l}\,  H^T_l(-t_{l})}
\otimes  {\cal T}_{r\,} e^{- i\int^{t_{r}}_{0} dt_{r}\,  H_r(t_{r})}~ |\Psi\rangle\,,
\end{equation}
where the left/right  time parameters denoted as  $t_{l,r}$ may run independently in general  but
our boundary time $u$ is related to them by $t_r=-t_l =u$. Here, $H_{l/r}(u)$ is the Hamiltonian obtained from (\ref{lagrangian}) and  ${\cal T}_{l}/{\cal T}_{r}$ represents the time ordering in the direction where $-t_{l}/t_r$ increases, respectively.

For the undeformed case, % without any deformation, 
the Euclidean evolution operator is given by 
\be
U_0=e^{-\frac{\beta}{2} H_0} \,,
\ee 
and the thermofield double state becomes \cite{Maldacena:2001kr}
\begin{equation} \label{TFD0}
|\Psi\rangle_0 = \frac{1}{\sqrt{Z_0}}\sum_{n} e^{-\frac{\beta}{2} E_n} |\bar{n}\rangle_{l} |n\rangle_{r}\,,  \qquad  Z_0=\sum_{n} e^{-{\beta} E_n}\,,
\end{equation}
with an energy eigen-basis of $H_0$, 
 from which usual thermal correlation functions may be obtained as their expectation values. % of operators.

\begin{figure}[htbp]   
\begin{center}
\vskip0.1cm
\begin{tikzpicture}[scale=1.1]
\draw[blue, thick,-<](-2,0)--++(0,2);\draw[blue,thick] (-2,2)node[left]{\large$t_l$}--++(0,2);
\draw[thick,red, ->](2,0)--++(0,2);\draw[thick,red] (2,2)node[right]{\large$ t_r$}--++(0,2);
\draw[blue,decoration={coil,segment length=0.5mm,amplitude=0.15mm},decorate] (-2,2) arc (180:270:2);\draw[red,decoration={coil,segment length=0.5mm,amplitude=0.15mm},decorate] (2,2) arc (0:-90:2); \draw[fill=black] (0,0) circle (0.05cm);
%\draw[decoration={zigzag},decorate](-2,0)--(2,0); 
 \draw[thick,dotted](-2,2)--++(4,0);
%\draw[decoration={zigzag},decorate](-2,4)--++(4,0); 
\draw[decoration={zigzag,segment length=1mm,amplitude=0.5mm},decorate](-2,4)--++(0.2,+0.35); 
\draw[decoration={zigzag,segment length=1mm,amplitude=0.5mm},decorate](2,4)--++(-0.2,+0.35); 
\draw[decoration={zigzag,segment length=1mm,amplitude=0.5mm},decorate](-2,0)--++(0.2,-0.35); 
\draw[decoration={zigzag,segment length=1mm,amplitude=0.5mm},decorate](2,0)--++(-0.2,-0.35); 
%\draw[thick,brown,dotted] (-2,0)--++(4,4);
%\draw[thick,brown,dotted] (+2,0)--++(-4,4);
\draw (-2,0)--++(4,4);
\draw (+2,0)--++(-4,4);
\draw[blue,decoration={coil,segment length=0.5mm,amplitude=0.15mm},decorate] (6,4) arc (90:270:2);\draw[red,decoration={coil,segment length=0.5mm,amplitude=0.15mm},decorate] (6,4) arc (90:-90:2); \draw[fill=black] (6,0) circle (0.05cm);
 \draw[thick,dotted](4,2)--++(4,0);\draw[fill=black] (6,0) circle (0.05cm);
%\draw[decoration={zigzag},decorate](-2,4)--++(4,0); 
%\draw[blue,decoration={coil,segment length=0.5mm,amplitude=0.2mm},decorate] (4,2) arc (180:90:2);
%\draw[red,decoration={coil,segment length=0.5mm,amplitude=0.2mm},decorate] (8,2) arc (0:90:2); 
\draw[fill=black] (6,4) circle (0.05cm);
\end{tikzpicture}
\vskip-0.1cm
\caption{\small On the left, we depict  the lower half of the  Euclidean geometry combined with the subsequent Lorentzian
evolution. The former %of the Euclidean geometry 
is used to generate the deformed thermofield
initial state. The right figure illustrates the full Euclidean evolution which may be used to compute
the normalization factor $Z$ or the thermal expectation value (vev) of operators with an appropriate insertion of operator $O(t)$.
}
\label{EuctoLor}
\end{center}
\end{figure}
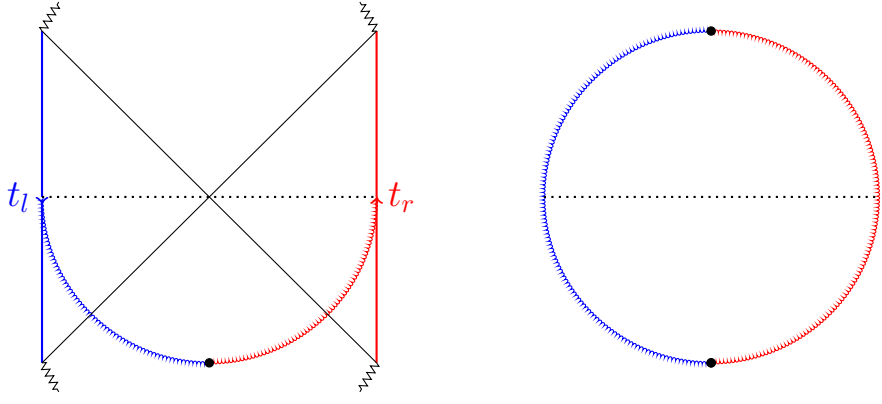

Now let us deal with the leading order of perturbation. In this case the Euclidean evolution $U$  consists of the left, the right
Euclidean evolution and a mid-point insertion of the operator for the vev deformation. Namely, it has a form of
\be
U= U_r\, e^\V \, U_l \,.
\ee
The left  and right evolution %evolution 
operators are basically obtained by an appropriate Euclidean continuation of the Lorentzian counterparts in (\ref{lorentzianevol}).  For the right side, we use the usual analytic continuation rule
with $t_r= -i t_E$; The right Lorentzian time ranged over $(-\infty,0)/(0,\infty) $ is mapped to the Euclidean time $t_E$
ranged over $(-\beta/4,0)/(0,\beta/4)$, respectively. On the right side of Figure \ref{EuctoLor}, we depict 
the Euclidean version of the (undeformed) bulk geometry where the blue/red colored semicircle is for the 
left/right boundary, respectively. The red-colored (right-side) semicircle is covered by the Euclidean time range 
$t_E \in (-\beta/4, \beta/4)$ where the full circle has a circumference $\beta$. The continuation of the left side is more subtle; We use an analytic continuation rule $t_l \rightarrow i t^l_E $ but with a further shift by
$\pm \frac{\beta}{2}i$ leading to $t_l= i t_E=i(t^l_E \pm \beta/2)$. Then the Lorentzian time $t_l$ %({\color{red} $t_{l}$???})   
ranged over $(-\infty,0)/(0,\infty) $ is first mapped to %the Euclidean time 
$t^l_E$
ranged, respectively, over $(-\beta/4,0)/(0,\beta/4)$ but, including the shift,  to the Euclidean time  $t_E$
ranged  over $(\,\beta/4,\,\beta/2)/(-\beta/2,-\beta/4)$, respectively. This range of the Euclidean time %that is for the left side
is depicted by the blue-colored semicircle on the right panel of  Figure \ref{EuctoLor}. Combining the left  and right semicircles, the full boundary circle
is covered by the range $(-\beta/2,\beta/2)$. With this preparation, it is proposed in \cite{Bak:2007qw} that the left  and right Euclidean evolutions are given, respectively, by
\be
%U_{l/r}={\cal T}_E e^{-\int_{\negthinspace-\negthinspace\frac{\beta}{2}/\negthinspace-\negthinspace\frac{\beta}{4}}^{\negthinspace-\negthinspace\frac{\beta}{4}/\,\,  0} dt_E \, H_{l/r}(-i t_E)}
U_{l}={\cal T}_E \, e^{-\int_{\negthinspace-\negthinspace\frac{\beta}{2}}^{\negthinspace-\negthinspace\frac{\beta}{4}} dt_E \, H_{l}(-i t_E)}
, \ \ \ U_{r}={\cal T}_E\,  e^{-\int_{\negthinspace-\negthinspace\frac{\beta}{4}}^{\,\,  0} dt_E \, H_{r}(-i t_E)}\,.
\ee
This proposal with ${\cal V}=0$ is tested for the   two-sided Janus black holes \cite{Bak:2007jm,Bak:2007qw} and shown to reproduce the expected vev function to the leading order precisely \cite{Bak:2007qw}; Indeed, one can also show that the vev for the 2d  two-sided Janus black hole %solution
in (\ref{ejanus})
  %\cite{Bak:2018txn}
is  reproduced by a similar computation, whose details will be omitted in this note. We shall not %further 
test this part of the proposal any further % in this note 
and, instead,  % Instead we shall 
focus on the vev deformation in the following.

On the left side of Figure \ref{EuctoLor}, we depict  the lower half of  Euclidean geometry combined with the subsequent Lorentzian
evolution where the former
is used to generate the deformed thermofield %double
initial state.
To prepare a thermofield initial state by the Hartle-Hawking construction, we need to patch the Euclidean part to the Lorentzian one along an appropriate hypersurface. To the leading order of deformation, the bulk geometry will not be deformed and one may patch the  geometries along the time-reversal symmetric slice at $\tau=0$.  %we may restrict ourselves to the time-reversal symmetric case around $\tau=0$. 
%in order to simplify the discussion.  
%The relevant patched Penrose diagram from Euclidean to Lorentzian spacetime  is given  in Figure~\ref{EuctoLor}. 

%

Let us now include the vev deformation and construct the corresponding mid-point inserted operator given by
\be  \label{Vdef}
\V=\frac{\beta}{2}\sum^\infty_{n=1} a_n \, e^{i n\tau^v_n} \, O_n \big( 
\text{\footnotesize $\frac{\beta}{4}$} i 
%\frac{\beta}{4}
\big)
\ee
with 
\be  \label{Ondef}
O_n (-it_E) = {\cal P}_{n-1} %\Big(-\frac{\beta}{2\pi}\frac{d}{d t_E} \Big) 
\, O (-it_E) \,,
\ee
where a differential operator ${\cal P}_{n-1} \, (n=1,2,\cdots)$  will be further specified below. The operator $e^{\cal V}$ may also be realized
as a Euclidean evolution operator 
\be
e^\V = {\cal T}_E\, e^{+\int^{\,\, \epsilon-\frac{\beta}{4}}_{-\epsilon-\frac{\beta}{4}} dt_E g(t_E) O(-it_E)}
\ee
with
\be 
g(t_E)=\text{\footnotesize $\frac{\beta}{2}\,$}\delta
\big(t_E+\text{\footnotesize$\frac{\beta}{4}$}\big)
\sum^\infty_{n=1} a_n e^{ i n \tau^v_n} \, \P_{n-1} \,,
\ee
where we take the $\epsilon \rightarrow 0$ limit in the end. 
In Figure \ref{EuctoLor}, these insertions %of operators
 are represented by the black dots in the lower or the upper half of the Euclidean evolution.

One may also introduce a Hamiltonian along the lower half of the Euclidean evolution by
\be 
H(-it_E) =\left[\begin{array}{lll}
\ H_l (-it_E)\,,  &  & \ \, \,-\frac{\beta}{2}\,\, < \, t_E < \negthinspace\negthinspace- \epsilon\negthinspace-\negthinspace \frac{\beta}{4}  \\
 -g(t_E) O(-it_E)\,,  &  &  - \epsilon\negthinspace-\negthinspace \frac{\beta}{4}\negthinspace <\, t_E <\, \, \epsilon\negthinspace-\negthinspace \frac{\beta}{4}\\
\ H_r (-it_E)\,,  &  &  \ \ \epsilon\negthinspace-\negthinspace \frac{\beta}{4} <\,  t_E \,< \, \, \, 0
\end{array}\right. \,.
\ee
Then $U$ will be given by
\be
 U= {\cal T}_E\, e^{-\int^{\, 0}_{\ng -\ng \frac{\beta}{2}} dt_E H(-it_E)}
\ee
including the contribution from the mid-point insertion. In the Euclidean space, the counterpart of the Lorentzian unitarity
requires the reflection positivity
\be
H^\dagger (-it_E) = H( i t_E)\,,
\ee
by which one may also introduce the Euclidean Hamiltonian $H(-i t_E)$ for the upper half of the full thermal circle ranged over $(0, \beta/2)$. Based on this, it is straightforward to show that
\be
U^\dagger = {\cal T}_E\, e^{-\int_{\, 0}^{\frac{\beta}{2}} dt_E H(-it_E)}
\ee
and the Euclidean evolution along the full-circle is then given by $U^\dagger U$.

We shall now come to the purely vev deformation without any source terms introduced. %completely. 
The Euclidean evolution operator $U$ in this case reads
\be
U=e^{-\frac{\beta}{4} H_0}\, e^{\cal V} e^{-\frac{\beta}{4} H_0} \,,
\ee
from which  we would like to reproduce (\ref{vevdef}) with (\ref{leadingvev}). Below we shall focus on $\langle O_r(t)\rangle$; %may treat   
Of course $\langle O_l(t)\rangle$ may be treated 
in the same way, %.
% However 
but
we shall not repeat the latter computation in this note.
Let us first note that\footnote{Similarly $\langle {O_l}(t) \rangle$ is given by $\langle\Psi |\, {O}(t)^T \otimes {\bf 1} |\Psi\rangle$. }
\begin{equation} \label{}
\langle {O_r}(t) \rangle=\langle\Psi | {\bf 1}\otimes { O}(t) |\Psi\rangle  
%(\langle {O_r}(t) \rangle=\langle\Psi |{ O^T}(t)  \otimes {\bf 1}|\Psi\rangle  )
\end{equation}
which may be evaluated perturbatively. 
%One of our goal in the following computation is to identify the value of $t_{E}$ in $g(t_{E})$ to obtain our initial thermofield double state. Then, let us consider the operator insertion of ${\cal O}$ in the right boundary at a Lorentzian time $t$ and then compute its vev in order to match its value with the bulk result. This procedure reveals that the value of $t_{E}$ in our setup is given by the mid points of the Euclidean time evolution,  $-\frac{1}{4}\beta$.  Concretely, the straightforward computation of the vev of ${\cal O}$ inserted at the right boundary  gives us 
%
%\begin{equation} \label{}
%\langle\Psi | {\bf 1}\otimes {\cal O}(t) |\Psi\rangle =\langle {\cal O} \rangle_{\beta_{L/R}} =  \frac{1}{Z}\text{tr}_{R}\Big[{\cal O}(t) U(0,{\textstyle -\frac{\beta_{L}}{2} }) U({\textstyle \frac{\beta_{R}}{2} },0) \Big]\,,
%\end{equation}
%
%Working in the leading order 
With the fact % condition 
that the undeformed vev of ${O}$ vanishes, % in the leading order, 
it is straightforward to show that
\be
\langle {O_r}(t) \rangle ={\cal A}_-(t) +{\cal A}_+(t) + {\cal O}(a_n^2) \,,
\ee
where
\begin{equation} \label{vevpert}
{\cal A}_\pm(t)=\lim_{\epsilon\rightarrow 0}   \int^{\,\, \epsilon\pm\frac{\beta}{4}}_{-\epsilon\pm\frac{\beta}{4}} dt_E\,  g(t_E) \, \frac{1}{Z_{0}}\text{tr} \Big[ e^{-\beta H_{0}}\,  { O}(t)\,{ O}(-it_E) \Big] \,. %+ {\pmb{\mathcal O}}(g^{3})\,,
\end{equation}
%
%\begin{equation} \label{vevpert}
%\langle\Psi | {\bf 1}\otimes {\cal O}(t) |\Psi\rangle =   \frac{1}{Z_{0}}\int^{\beta}_{0}ds~ g(s) \text{tr}_{R}\, \Big[ e^{-\beta H_{0}}\,  {\cal O}(t)\,{\cal O}(-is) \Big] + {\pmb{\mathcal O}}(g^{3})\,,
%\end{equation}
%
%where $H_{0}$ and $Z_{0}$ denote  the undeformed Hamiltonian and partition function, respectively. Here, we have denoted the undeformed inverse temperature as $\beta$, which is the zero-th order left/right temperature,  and    used $\beta$-periodicity to adjust the integration range in such a way that $0 \le s \le \beta$. 
For the boundary CFT, the thermal two-point correlation is well known as
\begin{equation} \label{integ}
 \frac{1}{Z_{0}} \text{tr} \Big[ e^{-\beta H_{0}}\,  {O}(t)\,{ O}(-i t_E) \Big] =\frac{\text{\footnotesize$2\pi/\beta^{2}$}}{1-\cosh \frac{2\pi}{\beta}({t}+i{t_E})}\,,  
\end{equation}
where its normalization is worked out in  \cite{Bak:2017rpp} (See also \cite{Bak:2017xla}).
%
%where we have used the normalized variables such as $\hat{t} \equiv \frac{2\pi}{\beta}t$ and $\hat{s} \equiv \frac{2\pi}{\beta}s$.   According to the AdS/CFT correspondence,   the leading order vev of the operator ${\cal O}_{1}$ dual to the bulk scalar mode $\cos(\tau-\tau_{1})$ can read from the bulk solution as 
%
%\begin{equation} \label{}
%\langle {\cal O}_{1} \rangle_{\beta} =  \frac{1}{2} \frac{2\pi}{\beta_{0}}\, \frac{a_{1}}{\cosh\hat{t}}\,  \cos(\tau-\tau_{1})+\cdots\,,
%\end{equation}
%
%where we used the relation between the bulk time $\tau$ and the boundary time $\hat{t}$ given by $\cos\tau = 1/\cosh\hat{t}$.
In order to reproduce (\ref{leadingvev})  from the above perturbative computation  with only $a_1$ turned on, we need to take  $g(t_E)$  in~\eqref{vevpert} as
\begin{equation} \label{g1}
g_{1}(t_E) =  \frac{\beta}{2}\Big[e^{i\tau^v_{1}}\, \delta (t_E + \beta/4 ) + e^{-i\tau^v_{1}}\, \delta (t_E - \beta/4 )  \Big] \,,
\end{equation}
where we have introduced a notation
$g(t_E)=\sum^\infty_{n=1} a_n g_n(t_E)$ and used the identity    (Recall that $\cosh \frac{2\pi}{\beta} t =1/\cos \tau$)
\be
\frac{1}{1-\cosh  \frac{2\pi}{\beta}\big(t- \frac{\beta}{4}i \big) }=e^{-i\tau}\cos \tau\,.
\ee
%This result also confirms that the mid-point insertion of the operator ${O}$  is of the Euclidean time evolution. 
From this, we conclude that $\P_0 =1$. % which is introduced in (\ref{Vdef}) and (\ref{Ondef}). 
One may also check that no  choice other than the  above insertion point  works in reproducing  (\ref{leadingvev}).

%To obtain the  operator  dual to higher modes in the bulk scalar, let us recall the bulk result for the higher modes. Using the expansion of $\sin$-function at the right boundary as
%
%\begin{equation} \label{}
%\sin n(\mu+ \frac{\pi}{2})\Big|_{\mu\rightarrow\frac{\pi}{2}} = (-1)^{n-1}n \cos\mu + %\pmb
%{O}\Big((\cos\mu)^{3}\Big)\,, 
%\end{equation}
%
%one can deduce that the vevs of boundary operators dual to higher modes for the bulk scalar field are given by
%
%
%\begin{equation} \label{}
%\langle {\cal O}_{n} \rangle_{\beta} =  \frac{1}{2} \frac{2\pi}{\beta_{0}}\, \frac{(-1)^{n-1}n\, a_{n}}{\cosh\hat{t}}\,  \cos n(\tau-\tau_{n})+\cdots\,,
%\end{equation}
%
For $g_n$ with $n  > 1$,  one could adopt the following recursive strategy. The integration in~\eqref{vevpert} together with the expression in~\eqref{integ} can be used to find the differential operator  $\P_{n}$ acting on a function of $t_E$ as %follows:
%
%
%\begin{align}    \label{}
%\int^{2\pi}_{0}d\hat{s}\,  \delta({\textstyle \hat{s} - \frac{3}{2}\pi})\,  P_{n}\cdot \frac{1}{1-\cosh(\hat{t}+i\hat{s})} &= P_{n}\cdot \frac{1}{1+i\sinh(\hat{t}+i\hat{s}')} \bigg|_{\hat{s}'=\hat{s} - \frac{3}{2}\pi=0} 
% \nonumber \\
%&= (-1)^{n}(n+1)~  e^{-i(n+1)\tau}\cos\tau\,,
 %  \nonumber 
%\end{align}
%
%
\be
 \P_{n}(x)%\Big(-\frac{\beta}{2\pi} \frac{d}{d t_E}\Big) 
\frac{1}{1-\cosh \frac{2\pi}{\beta}(t+i t_E)
} \bigg|_{t_E=-\beta/4} 
= (-1)^{n}(n+1)~  e^{-i(n+1)\tau}\cos\tau\,,
\ee
where $x$ denotes the differential operator $-\frac{\beta}{2\pi} \frac{d}{d t_E}$.  
%Within this equation, one may view that $x$ is equivalently acting on $t$ by $x=-i\frac{\beta}{2\pi} \frac{d}{d t}$.
By a change of variables $\cosh \frac{2\pi}{\beta} t =1/\cos \tau$, %with $z\equiv \hat{t} +i\hat{s}'$, one can define the differential operator $P_{n}$ by the condition of 
the above expression becomes
\begin{equation} \label{pnoperation}
\P_{n}(x) e^{-i\tau}\cos \tau  = (-1)^{n}(n+1)~ e^{-i(n+1)\tau}\cos \tau\,.
\end{equation}
with $x= -i\cos\tau \frac{d}{d\tau}$. % as a further change of variable.
Acting with the differential operator $-i\cos\tau \frac{d}{d\tau}$ on this defining equation  once more, one may obtain the following recursion relation %for $\P_{n}$
\begin{equation} \label{}
2x\,\P_{n}(x) = (n+1)\P_{n+1}(x) + (n-1)\P_{n-1}(x) 
%\,, \qquad  {\textstyle x= -i\cos\omega\, \frac{d}{d\omega} =-\frac{d}{d\hat{s}} = -\frac{d}{d\hat{s}'} }\,,
\end{equation}
with initial conditions
\begin{equation} \label{}
  \P_{0}(x)=1\,, \qquad \P_{1}(x) =2x\,.
\end{equation}
One may notice that the solutions to %polynomials satisfying 
this recursion relation are given by a special type of  the Meixner-Pollaczek polynomials~\cite{NIST}\,, 
\begin{equation} \label{}
\P_{n}(x) = \P^{\lambda=1}_{n}(x\,;\, {\textstyle \phi=\frac{\pi}{2} })\,,
\end{equation}
which can be defined through a hypergeometric function as 
\begin{equation} \label{}
\P^{\lambda}_{n}(x\,;\, \phi ) = \frac{(2\lambda)_{n}}{ n!} e^{in\phi} F(-n,\lambda + ix\,;\, 2\lambda\,|\, 1-e^{-2i\phi})\,.
\end{equation}
%
%As a result, one can see that by taking $g_{n}$ 
%
%\begin{equation} \label{}
%g_{n}(\hat{s} ) =  \frac{\pi}{2}\Big[e^{in\tau_{n}}\, P_{n}({\textstyle \frac{d}{d\hat{s}} })\,  \delta({\textstyle \hat{s} - \frac{3}{2}\pi }) + e^{-in\tau_{n}}\, P_{n}({\textstyle -\frac{d}{d\hat{s}} })\, \delta(\textstyle{ \hat{s} -\frac{\pi}{2}} ) \Big]\,,
%\end{equation}
%
%the resultant expression matches completely with  the bulk $n$-mode.  
%This also means that ${\cal O}_{n}$ is given in the form of  $P_{n}{\cal O}$, and the Euclidean evolution operator, returning to the original range $-\frac{\beta}{2}\le  \hat{s} \le 0$, is given by
%
%
%\begin{align}    \label{}
%& U(0,{\textstyle -\frac{\beta}{2}})  = e^{-\frac{\beta}{4}H_{0}}e^{\frac{\beta}{4}{\cal V} }   e^{-\frac{\beta}{4}H_{0}} + \cdots\,,  \nonumber \\
%& {\cal V} = \sum^\infty_{n=0}a_{n+1}e^{i(n+1)\tau_{n+1}}{\textstyle P_{n}(-\frac{d}{d\hat{s}}){\cal O}(-i\frac{\beta}{2\pi}\hat{s})}\Big|_{\hat{s}=-\frac{\pi}{2}}\,.
%  \nonumber 
%\end{align}
%
%
This completes the identification of the inserted operator $\V$ that reproduces the vev in
(\ref{leadingvev}). In the above construction of the thermofield double state, we have not included higher order corrections as we mentioned repeatedly. For instance, consider the left-right asymmetric black hole spacetime, which arises in quadratic order of the above deformations generically. %, %it is naturally expected that 
In this case, the above 
mid-point insertion will not be %is no longer 
working anymore and, then, one needs another prescription % will be required, 
which, unfortunately,    we do not know how to arrange. %engineer. %deal with.
% how to proceed %further and %We 
We leave this issue  to  the future study. 

\section{SL(2,\bR) representation in the operator space}\label{sec5}
In the previous section, we have constructed the mid-point inserted operator $\cal V$ which is realized as a linear combination of the operators $\{ O_1, O_2, \cdots \}$. By each of this insertion, one may obtain the corresponding initial state $|\psi \rangle_{\text{\tiny $\cal V$}}$. This amounts 
to %a form of %
%nothing but 
a variation %example 
of
%an 
operator-state maps in general CFTs. 
%Namely for a given operator $\cal V$, the corresponding TFD initial state
This realization of states is highly nonlinear in terms of the coefficients
$\{a_1e^{i\tau_1^v}, a_2e^{2i\tau_2^v},\cdots \}$ especially including the gravity correction via the deformation of the dilaton field which is quadratic in $a_n$ as shown in (\ref{fullsoln}). 

Since our AdS$_2$ dynamics involves SL(2,\bR) symmetries in general, it is expected that
the inserted operators also transform  %transforming 
under the symmetries. 
% and form their representation. 
In this section, we would like to clarify how these operators $O_n$ form a 
representation of the SL(2,\bR) algebra
\be \label{sl2ralgebra}
[B, E]=iP\,, \ \ \ [E, P]=iB\,, \  \ \  [B,P]=iE \,,
\ee
where $B$, $P$ and $E$ denote the three SL(2,\bR) generators. For later purpose, let us also introduce a Casimir operator ${\cal C}=B^2+P^2-E^2=K_+ K_- -E(E-1)$ where the raising and lowering operators $K_\pm$ are defined by $K_\pm=B\mp i P$.
 We note that  this  realization of the symmetries  is already explored in %Ref.~
%\cite{Maldacena:2018lmt}, 
Section {\color{blue} 4.2} of Ref.~\cite{Lin:2019qwu},
where the generators of SL(2,\bR) are identified as\footnote{These generators are   
related to the ones in Ref.~\cite{Lin:2019qwu} by an automorphism $\tau \rightarrow \tau+\pi$.   } 
\begin{equation} \label{slgenerator}
B=- i\cos\tau \partial_{\tau}  \,, \qquad P =- i\Big(\sin\tau\partial_{\tau} + \frac{1}{\cos\tau}\Big)\,, \qquad E = i\Big(\partial_{\tau} + \tan\tau\Big)\,. 
\end{equation}
%
%Indeed, one can 
Below, we shall show how these generators are acting upon the space of $\P_n$ explicitly.
%From the above explicit construction of ${O}_{n}$ dual to higher modes of the scalar fields, one can check that these operators and their corresponding modes form an explicit representation of SL(2,R). To see this, one may note that the symmetry realization in AdS$_2$ space %on the Euclidean operator insertion 
%is already explored 
%Below we shall explicitly
%check that these operators act on %normalized 
%$\P_{n}(x)$'s as SL(2,R) generators. %satisfying the SL(2,R) algebra
%\be \label{sl2ralgebra}
%[B, E]=iP\,, \ \ \ [E, P]=iB\,, \  \ \  [B,P]=iE \,. 
%\ee
 % as will be explicitly checked below.  
%Instead of $B, P, {E}$ acting on $\P_{n}(-x)$, let us consider operators acting on $\P_{n}(x)$, which are related to the above ones by an automorphism. These operators are given by
%
%\begin{equation} \label{}
% B'= -i\cos\tau \partial_{\tau}  \,, \qquad  P' = -i\Big(\sin\tau\partial_{\tau} + \frac{1}{\cos\tau}\Big)\,, \qquad E' = i\Big(\partial_{\tau} + \tan\tau\Big)\,. 
%\end{equation}
%
First, using the orthogonality property of Meixner-Pollaczek polynomials given by~\cite{NIST}
\begin{align}    \label{}
\int^{\infty}_{-\infty}dx\, P^{\lambda}_{n}(x)P^{\lambda}_{m}(x)w(x\,;\lambda,\phi) & =   \frac{2\pi\,\text{\footnotesize $\Gamma(n+2\lambda)$}}{\,\,n!\,(2\sin\phi)^{2\lambda}}\, \delta_{nm}\,,  \quad  \lambda >0\,,  \  0<\phi<\pi\,, \nonumber
\end{align}
with  % \nonumber \\
$w(x\,;\lambda,\phi)=|\Gamma(\lambda\negthinspace +\negthinspace ix)|^{2}e^{(2\phi-\pi)x}$,
one may introduce a new set of orthonormal %ortho-normalized 
%Meixner-Pollaczek 
polynomials $e_{m}(x)$ defined by
\begin{equation} \label{}
e_{m}(x) \equiv  \sqrt{\text{\footnotesize $\frac{2}{m\pi }$}} \, \P_{m-1}(x)\,, \qquad m=1,2,\cdots\,.
\end{equation}
These normalized polynomials $e_{m}(x)$ satisfy then the following recursion relation 
\begin{equation} \label{eRecRel}
x\, e_{m}(x) = {\textstyle \frac{\sqrt{m(m+1)}}{2}\, e_{m+1}(x) + \frac{\sqrt{m(m-1)}}{2}\, e_{m-1}(x) }\,.
\end{equation}
This also implies that 
\begin{align}    \label{sl2repA}
B\cdot e_{m}\cdot e^{-i\tau}\cos\tau &= \left[  {\textstyle \frac{\sqrt{m(m+1)}}{2}\, e_{m+1}  + \frac{\sqrt{m(m-1)}}{2}\, e_{m-1} }\right] \cdot e^{-i\tau}\cos\tau  \,,
  \nonumber \\
P\cdot e_{m}\cdot e^{-i\tau}\cos\tau &= i\Big[ {\textstyle \frac{\sqrt{m(m+1)}}{2}\, e_{m+1}  - \frac{\sqrt{m(m-1)}}{2}\, e_{m-1} }\Big]\cdot e^{-i\tau}\cos\tau  \,,
  \nonumber  \\
E\cdot e_{m}\cdot e^{-i\tau}\cos\tau &= m\, e_{m}  \cdot e^{-i\tau}\cos\tau\,,
\end{align}
where the actions of $P$ and $E$ are computed using (\ref{pnoperation}) followed by  the $\tau$-space operations given by (\ref{slgenerator}). Thus these latter two computations require an auxiliary 
function $e^{-i\tau}\cos\tau$ on which $e_m$ is acting upon.
%Now, one can see that $e_{n}$'s  form a  representation of SL(2,R) known as the discrete series representation, $D^{+}_{j=1}$. 

% as follows. One 
 In fact, one may construct %a realization of 
the SL(2,\bR) generators that are  acting  on $e_{m}(x)$ directly. % , 
% including some operations of %with help of 
%finite translations  in %of  the argument 
%$x$, 
This may be achieved %We realize them 
by additionally including a finite translation operation given by %an exponentiation of the differential operator $\frac{d}{dx}$ %as 
(See~\cite{Araaya,Vilenkin} for mathematical precedents) %results)
\begin{equation} \label{}
e^{\pm i\frac{d}{dx}}f(x) = f(x \pm i)\,. % \qquad 
\end{equation}
%
%one can realize 
Then the three generators  may be realized as %$B',P',E'$ operators 
%as 
%
\begin{equation} \label{slxspace}
B= x\,, \ \  \  P = i\cos({\textstyle \frac{d}{dx}})\, x\,, %=i\Big[x\cos({\textstyle \frac{d}{dx}}) -  \sin({\textstyle \frac{d}{dx}})\Big]\,, 
\ \  \  E = \sin({\textstyle \frac{d}{dx}})\, x\,, %= x\sin({\textstyle \frac{d}{dx}})+\cos({\textstyle \frac{d}{dx}})  \,.
\end{equation}
which act on an arbitrary square-integrable complex function $f(x)$. 
%
%It is straightforward to check that $e_{m}(x)$ form a $D^{+}_{j=1}$ representation, indeed. 
To verify these expressions, we first note that $\{e_1(x), e_2(x), \cdots \}$ forms an orthonormal basis for any square-integrable function $f(x)$. The identification of $B=x$ is already introduced  in the above construction. 
%Since the action of $B$ %on $e_{m}(x)$ 
%is just a  multiplication by $x$, its action on $e_{m}(x)$  is given simply by~\eqref{eRecRel}.  
The expression for $E$ can be found as follows; One starts from the generating function of  $\P_{n}(x)$
given by
%Using the generating function of $\P_{n}(x)$ polynomials given by
%
\begin{equation} \label{}
G(t,x) \equiv \sum_{n=0}^{\infty}t^{n}\P_{n}(x) = \frac{1}{1+t^{2}}\, e^{2x\arctan t}\,.
\end{equation}
By integration of 
the both sides with respect to $t$ followed by a multiplication of $x$, one obtains
\be
\sum_{m=1}^{\infty}\frac{1}{m}t^{m}\, x \, \P_{m-1}(x) = \frac{1}{2}\, \left( e^{2x\arctan t}-1\right)
\ee
and, by a further action of $\sin({\textstyle \frac{d}{dx}})$ on  both sides,  is led to
\bea
&&\sum_{m=1}^{\infty}\frac{1}{m}t^{m}\,\sin({\textstyle \frac{d}{dx}}) \big[\, x \, \P_{m-1}(x)\big] = \frac{1}{2} e^{2x\arctan t} \sin 2\arctan t \nonumber\\
&& =\frac{t}{1+t^2} e^{2x\arctan t}  =\sum_{m=1}^{\infty}t^{m}\P_{m-1}(x)\,.
\eea
Thus, with $E=\sin({\textstyle \frac{d}{dx}})x$, $E \, e_m(x)= m \, e_m(x)$ which is the desired result reproducing the third line of (\ref{sl2repA}). This demonstrates our expression for $E$ in 
(\ref{slxspace}).
%one can check that 
%
%\begin{equation} \label{}
%E \cdot e_{n}(x) = \sqrt{\frac{2}{\pi n}}\frac{1}{2i}\Big[(x+i)\P_{n-1}(x+i) - (x-i)\P_{n-1}(x-i)\Big] = n\, e_{n}(x)\,,
%\end{equation}
%
%and the action of $P'$ can also be checked similarly. 
Finally, the expression for $P$ in (\ref{slxspace}) can be found from $-i[B,E]$, which is acting on 
$e_m$ by
\be
P\, e_{m} = i\Big[ {\textstyle \frac{\sqrt{m(m+1)}}{2}\, e_{m+1}  - \frac{\sqrt{m(m-1)}}{2}\, e_{m-1} }\Big]\,.\ee
It is also straightforward to check the generators constructed in this way satisfy the SL(2,\bR) algebra in (\ref{sl2ralgebra}) with ${\cal C}=0$. Among  unitary representations of SL(2,\bR), there is the so called discrete representation ${\cal D}^+_j$ (See for example \cite{Bak:2002rq}), which is realized in the Hilbert space 
\be 
{\cal D}^+_j = \{ |jm\rangle  ; m=j, j+1,j+2,\cdots\}
\ee
with $j$ real and positive, $K_- |jj\rangle=0$,  and ${\cal C}=-j(j-1)$. Now one may straightforwardly confirm that 
the above representation belongs to ${\cal D}^+_{j=1}$ with ${\cal C}=0$.

It is natural to anticipate  that the dimension $\Delta$ operator dual to  the massive scalar field   is represented by ${\cal D}^{+}_{j=\Delta}$ with ${\cal C}= - \Delta(\Delta-1)= -m^2$. It may be interesting to construct all the representations of operators dual to the massive scalar fields by our method. For a more group-theoretic approach to this topic, one may refer to the reference~\cite{Kitaev:2018wpr}. 
% Also in the above construction of the thermofield double state, we have not included higher order corrections as we mentioned repeatedly.  
%We leave this issue to the future study. 
%This requires further studies.

%Before going to  next sections, we would like to give some comments.
%Through the conventional perturbative AdS/CFT dictionary in our case, one may expect that the Fourier coefficients of our bulk deformations given by the dilaton $\phi$ can be determined completely by state vevs and source values on a boundary\footnote{The left or right single boundary information is sufficient to   determine  the coefficients since our bulk deformations may be extended the expression of the $2\pi$ periodicity in the global time coordinate.}.  
%As far as those coefficients are very small, this indicates the complete bulk reconstruction. Then,  one may conclude that a python's lunch cannot be allowed since the bullk reconstruction would be quite difficult if the python's lunch exists. On the contrary, as will be discussed in the next sections, classical python's lunch configuration is closely related to the formation of a large  behind-horizon wedge region separating the left and right Rindler wedges, which corresponds to large values of Fourier coefficients of the bulk solutions. This large value case is our main interest in the following sections, which is beyond the scope in applying the perturbative AdS/CFT dictionary.

\section{Conclusions  %($m^2=0$) 
}\label{sec9}
In this work, we have considered the most general normalizable and nonnormalizable  bulk 
deformations in Jackiw-Teitelboim gravity with a massless field which
corresponds to either vev deformations or source deformations of 
the thermofield double state 
in the dual boundary theory. 
Such deformations are, in general, left-right asymmetric, resulting 
in different black hole temperatures. Moreover, we have argued that, classically, 
the bulk profiles may not be fully recovered from the data collected
along the boundary cutoff trajectories. %In other words, 
Then the bulk
seems to contain more information than the cutoff boundaries and this might be responsible for
the behind-horizon degrees of freedom such as those of Python's lunches.

The deformed state can be prepared by inserting operators on the boundary 
of Euclidean AdS$_2$ in the context
of Hartle-Hawking construction. In the limit of small vev deformations,
we have explicitly identified the 
operators for the bulk deformations which are all inserted at the 
mid-point during the Euclidean time evolution along the lower half
of the boundary of the thermal disk. We have found that inserted operators
form a discrete SL(2,{\bf R}) representation $\mathcal{D}_{j=1}^+$ with
vanishing Casimir. Since the boundary system has SL(2,{\bf R}) symmetries,
it is natural to anticipate such a realization in the operator space.
If the matter is massive with mass $m$ instead of massless, the corresponding
SL(2,{\bf R}) representation would be $\mathcal{D}_{j=\Delta}^+$ with
$\mathcal{C} = -m^2$. See~\cite{Kitaev:2018wpr} for this conclusion from a slightly different perspective. 

%It would be interesting to investigate more precisely the origin of SL(2, {\bf R}) symmetries in the space of inserted operators. 

In constructing the operators corresponding to vev functions, we have ignored
higher order corrections. The approximation allows us to work in
left-right symmetric undeformed geometries, since the asymmetry arises  
 starting from the second order in deformations. Inclusion of higher order terms
would involve asymmetric black hole spacetime, and the mid-point insertion of
operators would no longer be a valid prescription. We leave this issue to 
the future study. Finally, it would be interesting to generalize the 
results of this paper by turning on both the vev and the source 
deformations at the same time, for which black holes are further excited
as discussed in~\cite{Bak:2021qbo}.

\subsection*{Acknowledgement}
%We thank Juan Maldacena for enlightening discussions for 
%clarifying the nature of interactions in traversable wormhole systems.
We would like to thank Andreas Gustavsson for careful reading of the manuscript. 
DB was
supported by
%NRF Grant 2020R1A2B5B01001473, by  Basic Science Research Program
%through NRF %National Research Foundation 
%funded by the Ministry of Education
%(2018R1A6A1A06024977). 
the 2023 Research Fund of the University of Seoul.
 C.K.\ was supported by NRF Grant 2022R1F1A1074051.
S.-H.Y. was supported by 
%the National Research Foundation of Korea(NRF) 
NRF Grant  %with the grant number NRF- 
2021R1A2C1003644 and supported  by Basic Science Research Program through the NRF funded by the Ministry of Education (NRF-2020R1A6A1A03047877).

\end{document}